\documentclass[12pt,preprint]{aastex}
\usepackage{natbib}
\usepackage{amsmath}
\begin{document}
\newcommand{\crc}{cells$/r_c$}
\newcommand{\rr}[1]{$R_{#1}$}

\newcommand{\yfc}{\cite{yirak2009}}

\newcommand{\jet}{-1}
\newcommand{\control}{0}
\newcommand{\controlemiss}{0E}
\newcommand{\allcrush}{2}
\newcommand{\allcross}{1}
\newcommand{\bigdisp}{4}
\newcommand{\smalldisp}{3}
\newcommand{\clumps}{5}

\shorttitle{Clumpy Jets Parameter Study}
\shortauthors{Yirak et al.}

\title{The Evolution of Hetergeneous ``Clumpy Jets'': A Parameter Study}

\author{Kristopher Yirak\altaffilmark{1}, Ed Schroeder\altaffilmark{1}, Adam Frank\altaffilmark{1}, Andrew J. Cunningham\altaffilmark{2}}
\altaffiltext{1}{Department of Physics and Astronomy, University of Rochester, \
Rochester, NY 14620 \\Email contact: yirak@pas.rochester.edu}
\altaffiltext{2}{Lawrence Livermore National Laboratory, Livermore, CA 94551}

\begin{abstract}
We investigate the role discrete clumps embedded in an astrophysical jet play on the jet's morphology and line emission characteristics. By varying clumps' size, density, position, and velocity, we cover a range of parameter space motivated by observations of objects such as the Herbig Haro object HH~34. We here extend the results presented in \yfc{}, including how analysis of individual observations may lead to spurious sinusoidal variation whose parameters vary widely over time, owing chiefly to interacts between clumps. The goodness of the fits, while poor in all simulations, are best when clump-clump collisions are minimal. Our results indicate that a large velocity dispersion leads to a clump-clump collision-dominated flow which disrupts the jet beam. Finally, we present synthetic emission images of H-$\alpha$ and [SII] and note an excess of [SII] emission along the jet length as compared to observations. This suggests that observed beams undergo earlier processing, if they are present at all.
\end{abstract}

\section{Introduction}
Herbig-Haro (HH) objects are notable both for their high degree of collimation and their apparent ``knotty'' structure, as evident in observations, e.g. \cite{bally2002hh1/2}. Knots are typically axially-aligned emission spots bright in H-$\alpha$ and [SII], thought to be a result of shock heating \citep[][see e.g.]{hartigan2007bfields}. Spacing between knots is usually of order the size of the knots themselves, though larger voids where little to no emission is present may be observed. The origin of the knotted structure remains under debate. Several models have been proposed since their discovery \citep[][]{herbig1951, haro1952}; an overview may be found in \cite{reipurth2001ar}.

In \yfc{}, we proposed a generalization of the ``interstellar bullet'' model of \cite{norman1979}. Their model assumed a single, massive ``bullet'' ejected from a young stellar object (YSO) source, moving through the interstellar medium (ISM). The model in \yfc{} assumed the launching velocity and density profile of a HH jet to vary on a scale less than the jet radius, producing a ``clumped'' jet. The model produces morphology similar both to observations and to jets seen in the lab setting, see e.g. \cite{ciardi2008}. The clumped jet model achieves complex structures reminiscent of that seen in observations, including nonaxial, co-moving bow shocks achieved without precession. Our fully three-dimensional (3D) simulations utilize adaptive mesh refinement (AMR) in order to resolve clumps in the jet beam at sufficient resolution.

In this paper we extend the results presented in \yfc{} through a parameter space search designed to illuminate the role clump velocity dispersion, size, and density contrast play in the overall morphology. In \S~\ref{problem} we describe the model and our numerical methods. \S~\ref{results} highlights the results from the investigation, which are discussed in more quantitative detail in \S~\ref{analysis}. We conclude in \S~\ref{discussion}.

\section{Physical model and numerical scheme}\label{problem}

The model employed is discussed in \yfc; we briefly describe it here. Constant inflow conditions are imposed on the left ($x=0$) plane in a circle of radius $r_j=$, density $n_j$, temperature $T_j$ and velocity $v_j$. Starting at $t_0=8$ yr, at $\Delta t=8$ year intervals, spherical \emph{clumps} are placed within the jet beam near the $x=0$ plane, with radius, density, temperature, and velocity $n_c$, $T_c$, \& $v_c$, respectively. The values of these parameters are randomly chosen within certain bounds which differ depending on the simulation; see Table \ref{cj2_tab_parameters}. There were 7 cases in all, which are discussed in detail later in this section. The $y$, $z$ positions of the clumps also were randomly chosen, so long as a clump is placed entirely within the beam. In all cases, the ambient number density and temperature are $n_a=10$ cm$^{-3}$ and $T_a=2000$ K, respectively. In all cases except Case \clumps{}, the jet number density, temperature, and velocity are $n_j=100$ cm$^{-3}$, $T_j=2000$ K, and $v_j=150$ km s$^{-1}$ ($31.6$ AU yr$^{-1}$), respectively, producing an over dense ($\chi_{ja}\equiv n_j/n_a>1$), Mach 30 jet. The radius of the jet was $r_j=100$ AU. In Case \clumps{}, the jet number density was changed to be $n_j=1$ cm$^{-3}$. The simulations were performed using the \emph{ AstroBEAR}\footnote{More information about the $AstroBEAR$ code may be found online at http://www.pas.rochester.edu/$\sim$bearclaw/. A code-centered wiki is under active development, at http://clover.pas.rochester.edu/trac/astrobear} computational code \citep{cunningham2009}. In all cases except Case \controlemiss{}, the effects of radiative cooling is treated as a stiff source term based on the cooling curve of \cite{dalgarno1972}. Case \controlemiss{} includes a treatment of multiphysical dynamics, tracking helium and hydrogen ionization. Advected tracers are used to track the clump material throughout the simulations. 

\begin{table}[htbp]\centering\small
\begin{tabular}{lcccc}
\hline\hline
Case              & $r_c/r_j$ & $n_c/n_j$ & $v_c/v_j$ & Ionization?\\
\jet{}              & --    & --    & -- & No   \\
\control{}          & $0.25/0.50\pm0.15/0.75$ & 5 & $0.74/1.01\pm0.12/1.25$ & No\\
\controlemiss{}     & $0.25/0.50\pm0.15/0.75$ & 5 & $0.74/1.01\pm0.12/1.25$ & Yes\\
\allcrush{}         & $0.25/0.47\pm0.16/0.74$ & $9.3/21.4\pm11.1/44.2$ & $0.62/0.96\pm0.23/1.33$ & No\\
\allcross{}         & $0.25/0.52\pm0.15/0.74$ & $12.7/31.8\pm13.1/48.9$ & $0.91/1.03\pm0.08/1.25$ & No\\
\bigdisp{}          & $0.25/0.45\pm0.14/0.73$ & $5$ & $0.23/0.95\pm0.45/1.69$ & No\\
\smalldisp{}        & $0.25/0.45\pm0.14/0.73$ & $5$ & $0.90/0.99\pm0.06/1.08$ & No\\
\clumps{}           & $0.25/0.45\pm0.14/0.73$ & $5$ & $0.74/1.01\pm0.12/1.25$ & No\\
\hline
\end{tabular}
\caption{Physical parameters of the clumps instantiated in the simulations, normalized to the control jet values for radius, number density, and velocity of $r_j=100$ AU, $n_j=100$ cm$^{-3}$, and $v_j=150$ km s$^{-3}$, respectively. Values listed are minimums, means$\pm$standard deviation, and maximums for each case. \label{cj2_tab_parameters}}
\end{table}

In simulations where there will be many clump-clump interactions, the computational domain needs to be large enough that the resulting dynamics have time to evolve. Our computational domain has extents 4800 x 1200 x 1200 AU, or 48 x 12 x 12 $r_j$. This is larger than the domain in \yfc{}, allowing evolution beyond what was seen there. All simulations ran out to a time corresponding to the time it took the jet to cross the computational domain, roughly $t_{final}=200$ years. (Note that this crossing time implies a jet velocity of less than 150 km s$^{-1}$ if the simple formula $v_{bs}=v_j (1+\chi_{ja})^{-1/2}$ is employed, indicating that cooling plays an active role in the jet, removing energy from the system and slowing the jet slightly.)

The simulations had a base resolution of 160 x 40 x 40, with 3 additional levels of refinement in areas of interest---determined here by spatial fluid variable gradients---giving an effective resolution of 1280 x 320 x 320. This resolution thus gives 26 cells per jet radius; the size of the clumps is constrained such that the smallest clump is resolved by no less than 6 cells per clump radius, as in \yfc{}. Owing to the fairly low resolution in the clumps, we take the common step of employing a hyperbolic tangent smoothing profile which smoothes the outer 20\% of the clump in an effort to minimize grid-based seeding of instabilities. We note however that the conclusions drawn here concern the overall morphologies of the different cases and as such do not depend strongly on this aspect of the simulations.

The simulations are characterized by the characteristic time $t^*$, derived from considering the fate of a given clump as either dispersing or reaching the jet head. It makes use of the \emph{cloud-crushing time} given in \cite{klein1994} as approximately $t_{cc}=2 r \chi^{-1/2}/v$. The characteristic time is defined in \yfc:

\begin{equation}
t^* = \frac{2 r_c}{|\Delta v_c|} \sqrt{\chi_{cj}}\left(\frac{v_c}{v_j} \sqrt{1+\chi_{ja}^{-1}} - 1\right)\qquad (v_c\neq v_j)
\label{cj2_tstar}
\end{equation}
where $r_c$ is the radius of the clump $\Delta v_c$ its relative velocity to the jet velocity, $\chi_{cj}$ the density ratio of clump to jet, and $\chi_{ja}$ the density ratio of jet to preshocked ambient. If a clump is launched at a time $t_{launch}<t^*$, then---assuming no clump-clump interactions---it will reach the head of the jet before being dispersed. Conversely, if $t_{launch}>t^*$, the clump will be dispersed via hydrodynamic processing before travelling the length of the jet beam.

We wish to address three questions with these simulations: First, allowing clump density, size, and velocity all to vary, what are the effects of seeding clumps which all satisfy $t_{launch}<t^*$ versus seeding clumps which all satisfy $t_{launch}>t^*$? (We note that the results in \yfc{} had a mixture of clumps with $t_{launch}<t^*$ and $t_{launch}>t^*$.) Secondly, if we maintain clump density and clump size, what role does variation of clump velocity play on its own? Such a scenario may be motivated by supposing that the properties of the clumps being ejected from the source are all the same, but the ejection mechanism is unsteady, leading to different ejection velocities. Finally, if we effectively remove the background jet by making it under-dense, how does this compare to the other cases, both morphologically and in synthetic emission features?

Our suite of simulations is comprised of 7 different cases as detailed in Table~\ref{cj2_tab_parameters}. Case \control{} is meant to be a close reproduction of the simulation presented in \yfc{}---having a mix of crushing and crossing clumps---included and extended here for comparison. Cases \allcross{} \& \allcrush{} address the first question above: they possess clumps which vary in $r_c$, $v_c$, and $n_c$, with all the clumps satisfying $t_{launch}<t^*$ (``all cross'', Case \allcross{}) or all satisfying $t_{launch}>t^*$ (``all disperse'', Case \allcrush{}). Cases \smalldisp{} \& \bigdisp{} address the second question: if $n_c$ is held fixed, and the same sequence of $r_c$ values are chosen across the two simulations, what is the result of small velocity dispersion (Case \smalldisp{}) and large velocity dispersion (Case \bigdisp{})? Case \clumps{} addresses the final question: how well does the simulation reproduce observations when the jet is not present? Finally Case \jet{}, a steady jet without any instantiated clumps, is presented for comparison.

\section{Results}\label{results}

In what follows we refer to $upstream$ as being towards the launching plane (at $x=0$), and $downstream$ as away from the launching plane. When referring to the ``leading edge'' of the jet, we mean the position of jet and clump material which is located furthest downstream, i.e. the jet bow shock.

\begin{figure}[htbp]
\plotone{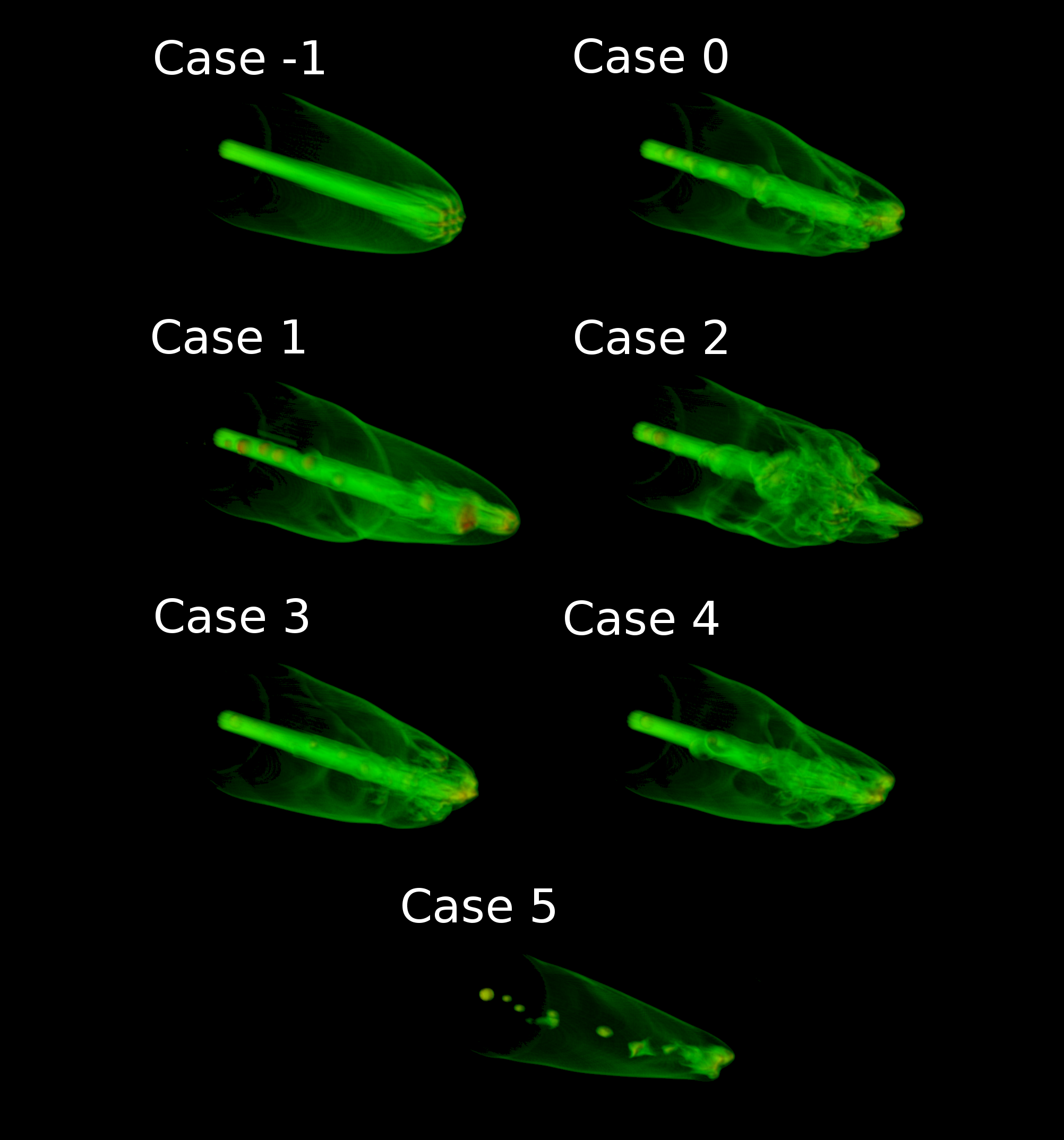}
\caption{Volumetric renderings of density for the 7 simulations at $t=120$ years, roughly two-thirds of the way through the simulation.\label{fig_vol}}
\end{figure}

Figure \ref{fig_vol} shows volumetric representations of density for each of the seven simulations at the same time, $t=120$ yr. Case \jet~exhibits the features commonly associated with a radiatively cooling, steady astrophysical jet, such as the relatively narrow width of the bow shock. Instabilities deriving from grid-based artifacts can be seen at the head of the jet, as is typical in Eulerian codes.

Case \control{}~appears as in \yfc{}. A mix of clump-crossing and clump-crushing events leads to a disrupted jet head as well as some clump-clump interactions along the jet beam. Overall, the morphology does not differ drastically from Case -1.

Case \allcross{}, in which the clumps are expected to cross the length of the jet, has an overall bow shock and jet beam similar to the jet-only case. The effects of the clumps in the jet beam are apparant near the leading edge of the jet, where their flattening into disc-like configurations increases internal pressure and serves to eject material laterally. The clumps' dispersal in the jet beam owing to $\Delta v\neq0$ is apparent in the shearing of material off the three clumps evident in this region, reminiscent of so-called \emph{spur shocks}, concave shocks located near the edge of the astrophysical jet beam that arc away from it \citep{heathcote1996hubble}. The dispersal of clumps in the jet beam naturally leads to such features.

The case in which the clumps are expected to be dispersed, Case \allcrush{}, exhibits much different morphology. Far downstream, the  jet beam becomes entirely disrupted as clump-clump interactions inject a large fraction of clump and jet material laterally into the surrounding jet bow shock and cocoon. The farthest-downstream feature of this simulation, rather than coming from the jet, is due to a particularly dense, fast-moving clump which has punctured the jet bow shock. The shape of the overall bow shock exhibits many small, filamentary features both parallel and perpendicular to the jet axis, and appears disrupted compared to Case \jet. There are small, distinct regions of enhanced density throughout. Thus, we see that the primary effect of choosing a host of clumps which satisfy $t_{launch}>t^*$ is not that they all disperse, but rather that it leads to a very high rate of clump-clump interactions, which is very disruptive to the overall flow. This may be characterized by the clump-clump collision rate, discussed in \S~\ref{analysis}.

Case \bigdisp{}, like Case \allcrush{}, displays disruption of the jet beam, though to lesser degree, due to the fact that the densities here are all $\chi_{cj}=5$, whereas in Case 1 they range up to $\chi_{cj}=44$. The location of maximum density enhancement is located near the head of the jet (terminal working surface), suggesting less lateral dispersal of material along the jet length. Nonetheless, the presence of the clumps does disrupt the jet bow shock, evident in apparent cavities (i.e., regions of lower density) when viewed at this angle.

Case \smalldisp{} also has a concentration of mass near the jet head, but does not show jet-beam disruption as in Case \bigdisp{}. While it, like Case \allcross{}, appears most similar to Case \jet{}, there are some differences. Case \smalldisp{} has more filamentary structure near the leading edge of the jet, and there are no major disruptions to the jet beam, whereas in Case \allcross{} the portion of the beam directly upstream the jet head is being disrupted.

Finally, Case \clumps{} noticeably differs from the others. While the overall bow shock shape is similar to that of Case \jet---smooth, no cavities or other disruptions---the rate of clump injection results in mostly isolated features along the jet axis. Near the leading edge of the jet, clumps merge to form a larger, dynamically evolving structure. Nearby upstream, a clump-clump collision results in a diamond-shaped object in this view. This feature is short lived and results in a local, small bow shock offset from the jet axis. Case \clumps~is of particular interest from a laboratory astrophysics point of view, as it is most similar to the results seen by e.g. \cite{ciardi2008} as discussed further in \S~\ref{discussion}.

\begin{figure}[htbp]
\plotone{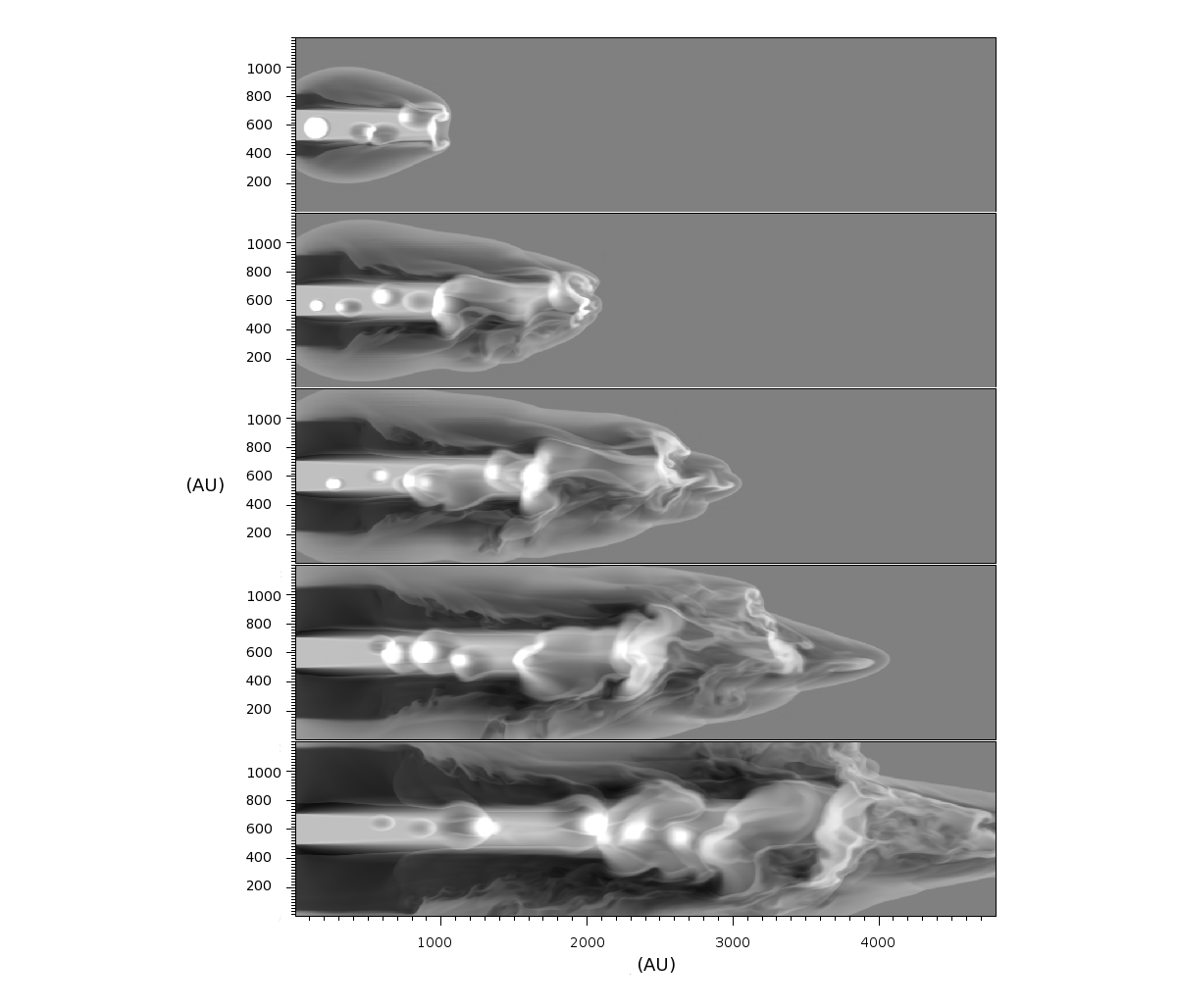}
\caption{Grayscale representation of density showing Case \allcrush~evolution at 5 different times: $t=40,\ 80,\ 120,\ 160,\ \&\ 200$ yr. \label{fig_evolution}}
\end{figure}

Next we consider the time evolution of the flows. Figure \ref{fig_evolution} shows a grayscale time evolution of Case \allcrush~at five different times throughout the simulation in the $x$, $y$ plane along the jet axis. The large velocity dispersions of the clumps are apparent in the first panel as individual forward- or backward-facing bow shocks, both of which create rarefaction regions in the jet beam. Such velocity dispersions, as discussed in \S~\ref{analysis}, result in rapid clump disruption and high rates of clump-clump interactions, further disrupting the jet beam. In later panels, as the simulation progresses, no coherent jet beam is evident beyond roughly $x=$1500 AU. Instead, increasingly complex multiple-shock interactions dominate the flow. This in turn produces little coherent structure from one panel to the next, with one notable exception. The large, slow-moving clump seen at $x=100$ AU in the first panel undergoes several clump-clump mergers. Its evolution results in the broad and increasingly diffuse disc-like structure seen at $x=1000,\ 1200,\ 2300,\ \&\ 3700$ AU, respectively, in the latter four panels.

\begin{figure}[htbp]
\plotone{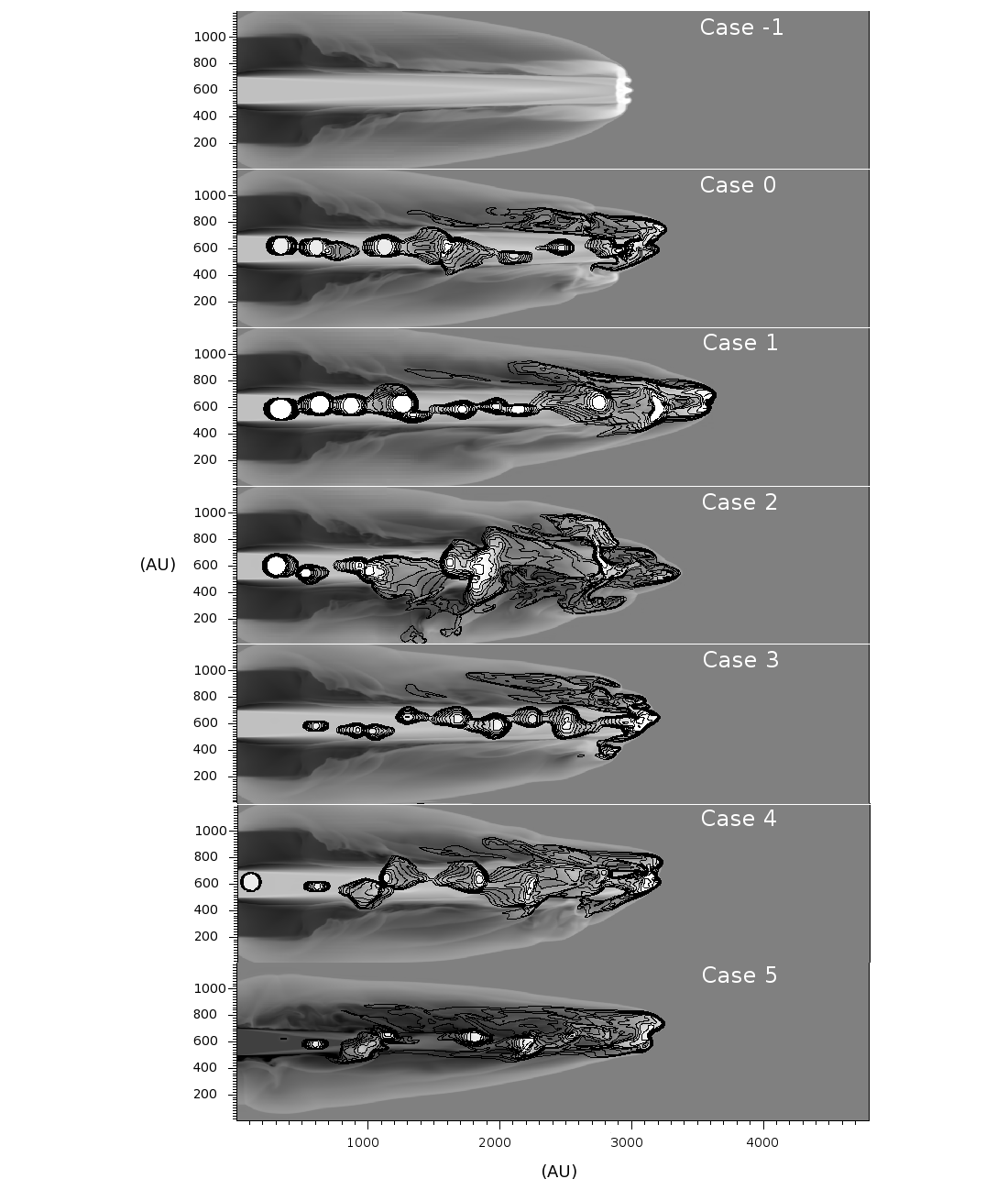}
\caption{All cases at $t=120$ years, grayscale slice along the axis of the jet with contours of clump tracer overlaid. Note that this time is the same as in Fig.~\ref{fig_vol}.\label{fig_compare120}}
\end{figure}

How does this evolution compare with the other cases? Figure \ref{fig_compare120} gives grayscale slices along the jet axis for each of the simulations at the same time, $t=120$ yr, which is the same as in Figure~\ref{fig_vol}. Overlaid are black contour lines, showing the distribution of clump material obtained via an advected tracer. We note significant clump material mixing in Case \allcrush, owing to the many clump-clump interactions. Even in relatively rarefied regions, such as $x\sim2300,\ y\sim400$, clump material is present. This suggests that the mixing of clump and jet material need not correlate directly with density. Further, were the chemical compositions of clump and jet material to differ, it would imply the emission signatures of these regions may differ depending on the amount of mixing.

\begin{figure}[htbp]
\plotone{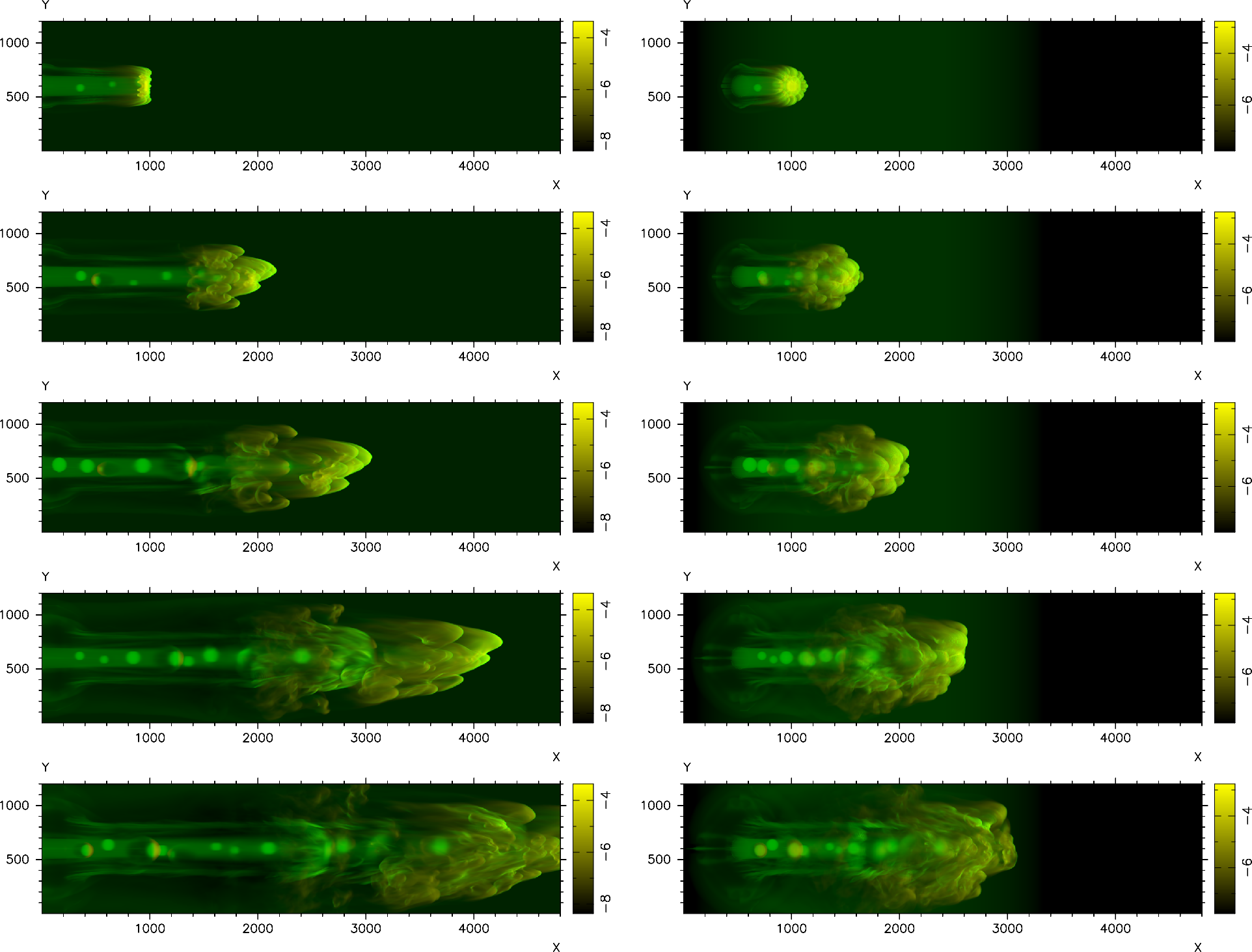}
\caption{Synthetic emission from Case 0E at same times as in Fig.~\ref{fig_evolution}. Images on the right are rotated 60$^o$ toward the viewer. Green represents H-$\alpha$ emission, and red [SII]. Multiple small bow shock features are apparent near the leading edge of the jet, reminiscent of HH object observations, while the jet beam has a higher ratio of H-$\alpha$ to [SII] than is typically observed. The units are logarithmic, summed emission intensity. The maximal unsummed intensity is of order $10^{-20}$ and $10^{-22}$ for H-$\alpha$ and [SII], respectively.\label{fig_emiss}}
\end{figure}

Finally, Figure \ref{fig_emiss} provides a line-of-sight integrated false color image of synthetic emission for Case \controlemiss, both with the jet axis located in the plane of the sky and rotated 60$^o$ toward the viewer. Red indicates [SII] emission, and green H-$\alpha$. Such a figure may be compared with similar images derived from \emph{Hubble Space Telescope} (HST) observations, see e.g. \cite{hartigan2005hh47}. These results are not intended for detailed comparison with any specific object, though we expect to carry forward simulations doing so in future papers. Here, we note that there is broad qualitative correspondence to observations in such features as multiple bow shocks near the head of the object. Although not presented here, there are short-timescale ($\sim$10 yr) variations in brightness as is observed in some HH objects. Projection effects also mimic those of observations; viewing the data set from multiple angles is necessary in order to correctly interpret the structure given only the synthetic emission. We do note that emission along the length of the jet indicates higher [SII] emission than is present in observations, perhaps supplying additional support for the removal of a coherent jet beam in astrophysical jet models.
\section{Analysis}\label{analysis}

\begin{figure}[htbp]
\plotone{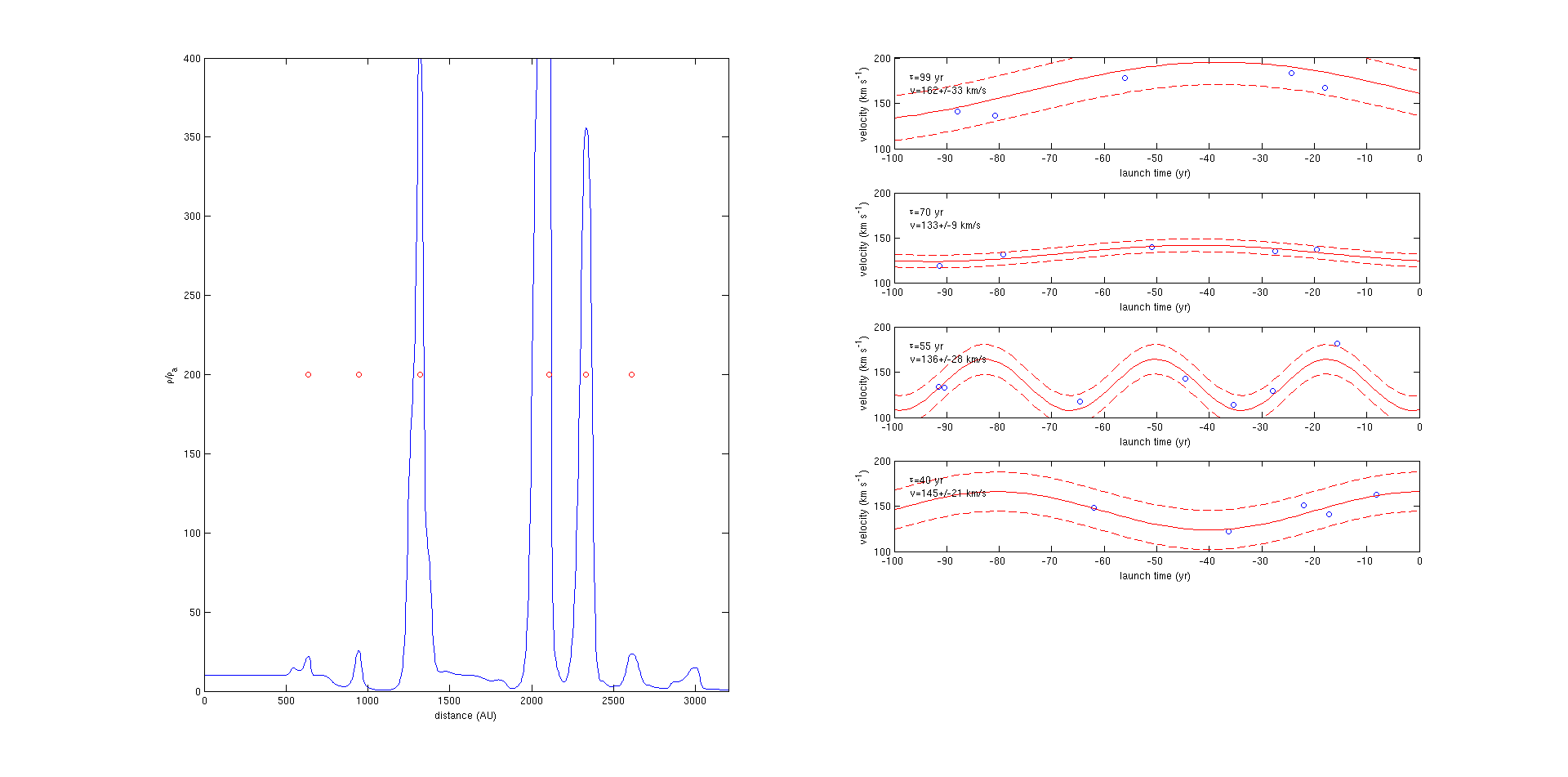}
\caption{Sinusoidal fitting of Case 2, as discussed in \S~\ref{analysis}. \label{fig_sine}}
\end{figure}

Of central importance in understanding HH objects and their central engines is the history of the launching. Being able to state if an HH object results from a steady outflow, or from an episodic one, would ultimately tie in to related processes occurring in the obscured launching region. One analysis method that has been employed is to extract position and velocity data from an observation, as in \cite{raga2002hh34/111}. In that work, the resulting data were fit with a two-mode sinusoid, and the resulting variables were used to perform axisymmetric hydrodynamic simulations. The simulations well matched the location of the leading bow shock and knots along the beam.

However, as noted in \yfc{}, this method may be erroneously applied to situations in which there is no sinusoidal launching, as is the case with the present model. \yfc{} found that sinusoidal fits were possible with a reasonable goodness of fit, but that the resulting period, in particular, was erroneous and varied widely as the simulation progressed.

We repeat and extend the sinusoidal analysis as discussed in \yfc{} for each case. Throughout the simulation, the locations and velocities of the clumps---defined as local density maxima along the jet axis---are used to fit a single sinusoid. The resulting amplitude, period, and goodness of fit are recorded. Figure~\ref{fig_sine} gives an example of the results for Case \allcrush, showing the result of fits at four times throughout the simulation. The left panel shows the axial density profile at $t=198$ yr, while the right panels show the fitting results for times $t=80,\ 110,\ 140, \&\ 198$ yr. As can be seen, the periods vary by more than a factor of 2, suggesting that the method is not robust. This variation is a direct result of clump-clump interactions, which create a constantly changing density profile in the jet beam, as discussed below.

\begin{figure}[htbp]
\plotone{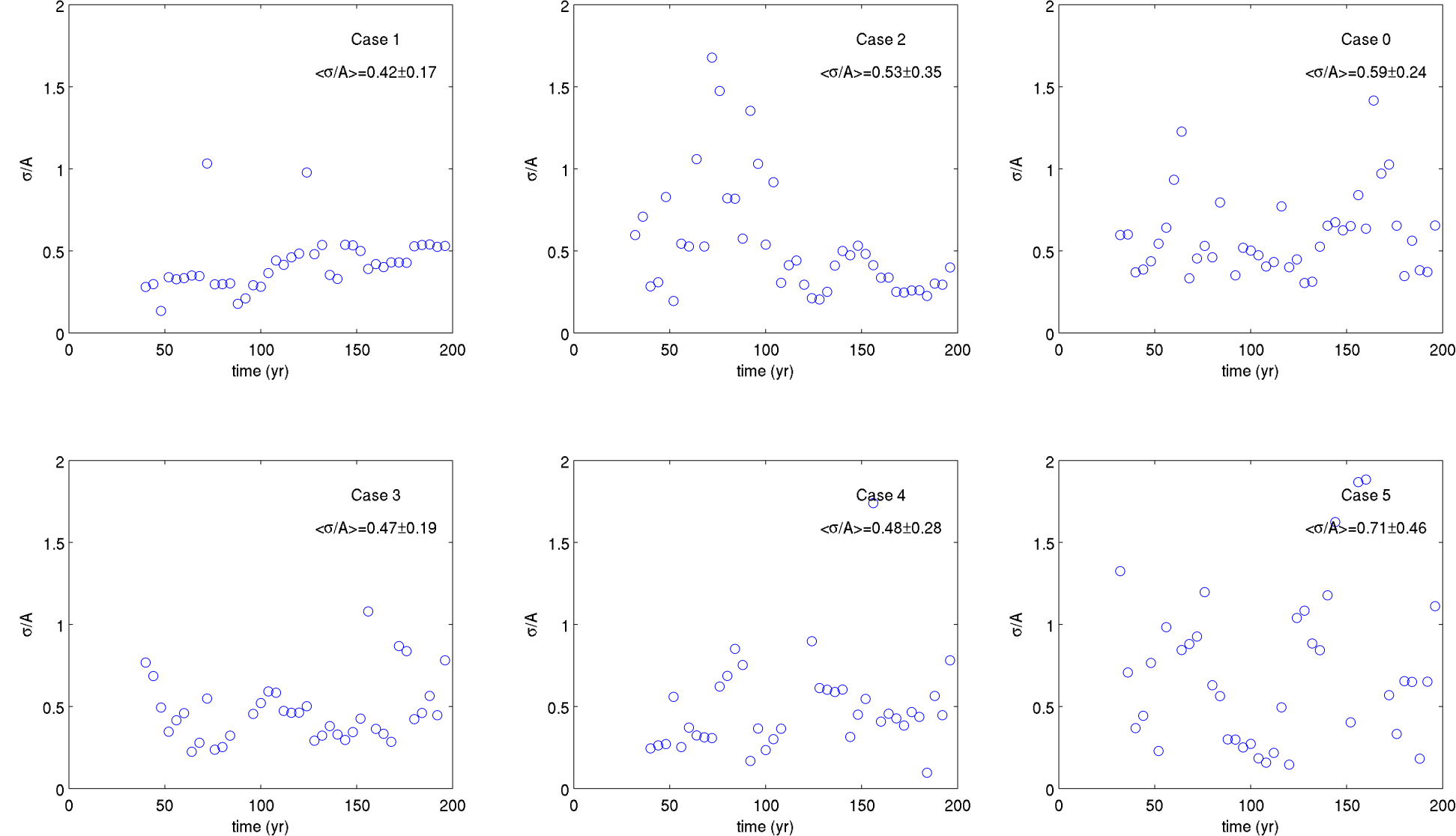}
\caption{Results of goodness-of-fit investigation of the sinusoidal fitting over time. The y-axis represents the 1-$\sigma$ confidence interval normalized by the sinusoidal amplitude at that time.\label{fig_gof}}
\end{figure}

How much should we trust these results? Figure \ref{fig_gof} presents a measurement relating to the goodness of the sinusoidal fit to each simulation, namely the 1-$\sigma$ confidence interval of the fit normalized by the sinusoidal amplitude at that time, denoted by $\sigma/A$. Also given are the time-averaged $\sigma/A$ values with their time-average standard deviations, where the time-averaged-quantity is denoted by $\langle\sigma/A\rangle$. Thus, values close to zero represent fits which matched closely the data, while those increasingly greater than zero were increasingly poor fits. Note that Case \jet{} is out of order, so that comparisons may be made more easily between similar cases (\allcross{} \& \smalldisp{}; \allcrush{} \& \bigdisp{}). The case with the most consistent and least standard deviation in goodness-of-fit was Case \allcross{}, though the large standard deviations of all cases implies this may not be statistically significant. Case \allcrush{}, which had the largest variation in both velocity and density, produced noticeably poorer fits than that of Case \bigdisp{}, in only the velocity variation was large while the densities were kept fixed. Finally, Case \clumps{} produced both poor goodness of fits as well as large variation from one time to the next. If, as we suppose, this case is the one most similar to the actual mechanism of YSO jets, this lends further credence to the idea that sinusoidal fits at a single instant in time will provide unreliable information regarding the history of the object.

Qualitatively, we see that consistency in periods and amplitudes, or reliability in goodness of fit, would be hampered by clump-clump collisions. One therefore may expect a negative correlation between clump collision rates and the goodness of the fitting results. However, it is possible that other factors in the simulations are of importance, such as the details of cooling or multidimensional effects. Are we therefore correct to assert the most important factor is the rate of collisions? We can address the importance of clump-clump interactions more quantitatively using a simplified model as follows. We can treat each clump's motion as linear and one-dimensional, clump $i$ having position $x_i(t)$ as
\begin{equation}
x_i(t) = v_i(t-t_{launch,i}) \qquad .
\end{equation}
We then require a model of the clump-clump interactions. We may take as a crude description of an interaction of clumps $i$ and $j$ as an inelastic collision at the moment that they touch, with the resulting clump having their combined mass ($M_{new}=4/3\ \pi \left(r_i^3n_i+r_j^3n_j\right)$), and density that is the average of $n_i$ and $n_j$, resulting in a new clump of radius $r_{new}=\left[3/4\ \pi^{-1} M_{new}/n_{new}\right]^{1/3}$. We treat the new clump's velocity by considering that as a clump is crushed over a time $t_{cc}$, its material becomes entrained in the bulk flow. This may be crudely approximated as $v_c$ asymptoting to $v_j$. Thus we take into account the dispersal of clump $i$ over a time $t_{cc,i}$ as a weighting factor when calculating the resulting merger velocity, as
\begin{equation}
v = \frac{W_i v_i + W_j v_j}{W_i + W_j}
\end{equation}
where $W_i=4/3\pi r_i^3 n_i (1-t/t_{cc,i})$. Table \ref{tab_linear} gives the results of the model using the clump parameters of Cases 0--5. The number of predicted mergers changes by a factor of two across simulations, with \allcrush{} \& \bigdisp{} having the most and \allcross{} \& \smalldisp{} the least. We find the number of mergers so predicted to agree well with what is observed in the simulations. Moreover, the location of the leading edge of the jet also agrees well with simulations. As before, the runs with few predicted mergers were those which were seen to have better goodness of fit results. Thus, we believe it correct to claim that clump-clump interactions are a very, if not the most, important process.

\begin{figure}[h!]
\centering
\plotone{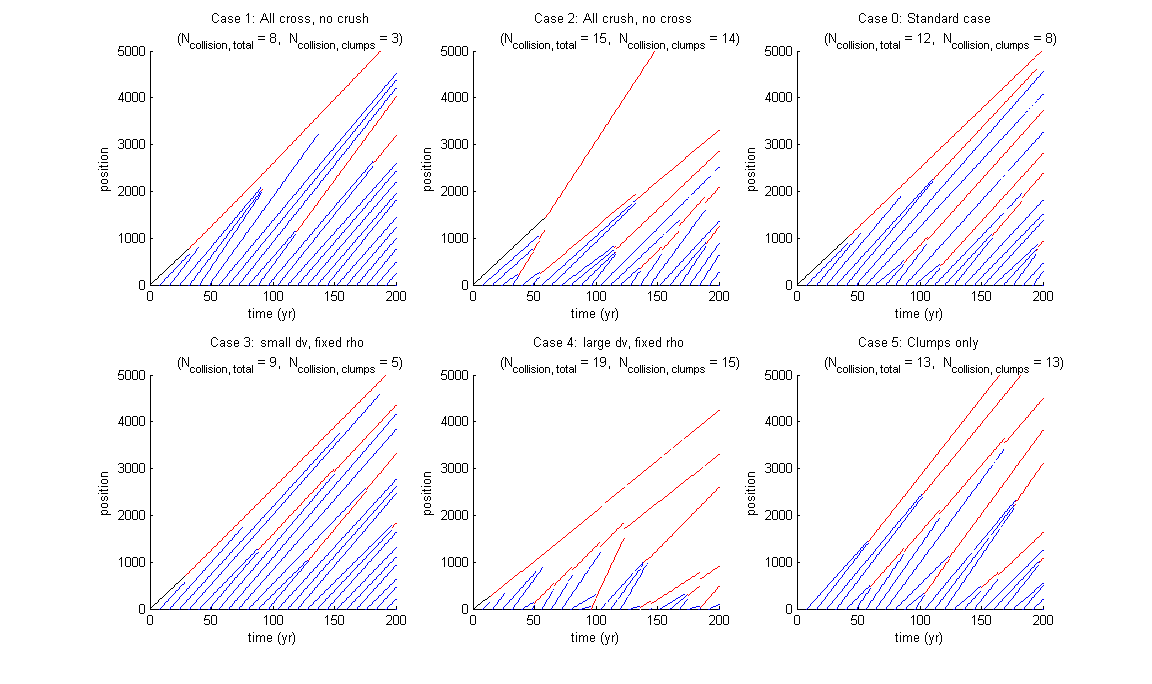}
\caption{Analytic predictions for number of clump collisions. $N_{col,total}$ count clump/clump collisions as well as clumps which overtake the jet bow shock. $N_{col, clumps}$ counts only clump/clump collisions. The results, including the expected position of the head of the jet, agree well with the simulations. Line colors change to red after a ``merger''.\label{cj2_collisions}}
\end{figure}

\begin{table}[htbp]\centering\small
\begin{tabular}{lccc}
\hline\hline
Case              & $N_{c,clump}$ & $N_{c,tot}$ & $<\sigma/A>$ \\
-1                & --           & -- & -- \\
0                 & 8  & 12 & $0.59\pm0.24$ \\
1                 & 3  & 8  & $0.42\pm0.17$ \\
2                 & 14 & 15 & $0.53\pm0.35$ \\
3                 & 5  & 9  & $0.47\pm0.19$ \\
4                 & 15 & 19 & $0.48\pm0.28$ \\
5                 & 13 & 13 & $0.71\pm0.46$ \\
\hline
\end{tabular}
\caption{Predicted number of clump-clump ($N_{c,clump}$) and clump-clump plus clump-bow shock (i.e., crossing events, $N_{c,tot}$) collisions for each case. The time-averaged quantity relating to goodness-of-fit of the simulation data from Fig.~\ref{fig_gof} are included for comparison. \label{tab_linear}}
\end{table}

\section{Discussion and Conclusion}\label{discussion}

In this paper, we have investigated the role that different parameters play in the clumped jet model introduced in \yfc{}. We found the jet bow shock to be affected in many of the cases, with the smoothest bow shock occurring in Case \allcross{} (excluding Case \jet{}). When the clumps' densities are kept fixed at $\chi_{cj}=5$, the role of velocity dispersion alone becomes apparent. With large dispersions as in Case \bigdisp{}, the jet beam is partially disrupted, with roughly clump-sized cavities appearing both in the jet beam and in the jet bow shock. Small dispersions, as in Case \smalldisp{}, on the other hand, do not affect the jet beam particularly but still have an effect on the jet bow shock.

What if the clump density is not held fixed? Instead, we may consider a joint constraint on velocity, size, and density based on Eq.~\ref{cj2_tstar}. In this case, the effects of the clumps are much more pronounced. In Case \allcrush{}, for example, which featured both high $\chi_{cj}$ and high velocity dispersion, the clumps demonstrated an ability to completely disrupt the jet beam, injecting both clump and jet material laterally into the surrounding jet cocoon (i.e., the rarefied region swept over by the laterally-expanding jet bow shock). In contrast to Case \smalldisp{}, even when the velocity dispersion is small as in Case \allcross{}, the increased range of densities results in a modified jet morphology. As discussed, this is due primarily to the clump-clump collisions, which in Case \allcross{} do not allow all clumps to cross the length of the jet but instead forces mergers, like those seen near the head of the jet in Fig.~\ref{fig_vol}.

However, a notable point is that, despite the variety of the cases mentioned above---and despite the jet beam disruption noted in Case \allcrush{}---the most disparate case is Case \clumps{}. With these parameters, the clumps are not numerous enough to mimic a jet beam. The fewer number of interactions results in an overall bow shock (analogous to the jet bow shock of the other cases) which is narrower or more slender. Thus, one could imagine that if propagating through a medium which itself is heterogeneous as discussed in \cite{yirak2008}, this case would see the least degree of energy transfer between the chain of knots and the surrounding heterogeneity. We should note that with the parameters as chosen, the injected mass (and therefore kinetic energy) of Case \clumps{} is substantially less than that in the other cases. It would therefore be instructive to extend these simulations with a focus on the injected momentum or kinetic energy.

Recent high energy density laboratory astrophysics (HEDLA) investigations provide an unique window into the behavior of fully 3-D radiative hypersonic MHD jets.  These experiments demonstrate that magnetized jet beams in the lab may rapidly break up into a sequence of quasi-periodic knots due to current driven instabilities \citep{ciardi2007labjet, ciardi2008}. These knots may be displaced from the nominal jet axis and may propagate with varying velocities. This results in morphologies qualitatively reminiscent of HH-jet beams. The present simulations do not employ magnetic fields; however, it remains an open question whether magnetic fields remain dynamically important on the length scales considered here (\cite{hartigan2007bfields, ostriker2001molecularclouds}). It seems plausible that a process similar to what's observed in the lab could occur in the astrophysical context, beginning with a beam close to the central engine which becomes disrupted owing to the kink and sausage instabilities on small to intermediate scales. This would result in a series of knots which continue to evolve as they propagate away from the central engine. Such a scenario would also explain the observed velocity differences between knots, attributable to the particulars of each knot's formation. The present simulations are an idealization of this model.

These simulations, as those in \yfc{}, feature on average a dozen computational zones per instantiated clump. As discussed in detail in \cite{yirak2010}, when radiative cooling is important this resolution may not sufficiently resolve the dynamics. However, as discussed in \cite{yirak2010} and references therein, higher resolution, leading to higher effective numerical Reynolds number in an inviscid code such as $AstroBEAR$, would allow faster mixing, but the gross morphology most likely would be little changed. The main effects we see in the simulations here have to do with the bulk mass of clumps impinging on their environment. The manner in which the subsequent mergers mix and intersperse certainly would be affected by higher resolution, and as such we have not attempted to quantify them at this stage. 

Finally, the simple linear model predicting the number of clump interactions would benefit and be improved by simulations of clump collisions. \cite{miniati1997}, among others, investigated the dynamics of clump-clump interactions in 2D axisymmetry without radiative cooling. They found, similar to the results in \cite{klein1994} of adiabatic clumps, that after collision the clumps disrupted and dispersed in a short time after collision. When \cite{miniati1997} include radiative losses, the resulting collisions result in coalesce, not unlike the mergers seen in the present work. It therefore would be fruitful to reexamine the existing model with these and similar results in mind, in order to synthesize a more sophisticated model of clumped astrophysical jets.

\acknowledgements
The authors wish to thank Jonathan Carroll and Brandon Shroyer for useful discussions. Support for this work was in part provided by by NASA through awards issued by JPL/Caltech through Spitzer program 20269 and 051080-001, the National Science Foundation through grants AST-0507519 as well as the Space Telescope Science Institute through grants HST-AR-10972, HST-AR-11250, HST-AR-11252. KY is a recipient of the Graduate Horton Fellowship provided by the University of Rochester Laboratory for Laser Energetics. We also acknowledge funds received through the DOE Cooperative Agreement No. DE-FC03-02NA00057.

\end{document}